# Natural and Intrinsic Vacancies in two-dimensional g-C$_3$N$_4$ for Trapping Isolated B and C Atoms as Color Centers

Manqi You[1,2], Chaoyu He[1,2*], Gencai Guo[1,2*], Jianxin Zhong[1,2*]

[1]*Institute for Quantum Science and Technology, Shanghai University, Shanghai 200444, China*

[2]*Hunan Key Laboratory for Micro-Nano Energy Materials and Devices, Laboratory for Quantum Engineering and Micro-Nano Energy Technology, and School of Physics and Optoelectronics, Xiangtan University, Hunan 411105, China.*

*hechaoyu@xtu.edu.cn; ggc@xtu.edu.cn; jxzhong@shu.edu.cn*

Color centers are vital for quantum information processing, but traditional ones often suffer from instability, difficulty in realization and precise control of locations. In contrast, natural intrinsic vacancy-based color centers in two-dimensional systems offer enhanced stability and tunability. In this work, we demonstrate that g-C$_3$N$_4$ with natural intrinsic vacancies is highly suitable for trapping B/C atoms to form stable color centers as qubits. With easily identifiable vacancies, B/C atoms are expectable to be placed at the vacancy sites in g-C$_3$N$_4$ through STM manipulation. The vacancy sites are confirmed as the most stable adsorption positions, and once atoms are adsorbed, they are protected by diffusion barriers from thermal diffusions. The most stable charge states are $C_V^{+2}/B_V^{+2}$, $C_V^{+1}/B_V^{+1}$ and $C_V^{0}/B_V^{0}$ in turn, with charge transition levels of 0.39 eV and 2.49 eV, respectively. Specifically, the defect levels and net spin of Cv/Bv can be adjusted by charge states. $C_V$, $C_V^{+1}$, $C_V^{+2}$, $B_V^{+1}$ and $B_V^{+2}$ exhibit optically allowable defect transition levels. The zero-phonon lines suggest fluorescence wavelengths fall within the mid-infrared band, ideal for qubit operations of stable initialization and read out. Furthermore, the Zero-field splitting (ZFS) parameter and the characteristic hyperfine tensor are provided as potential fingerprints for electron paramagnetic resonance (EPR) experiments.

Qubits, prized for their superposition and entanglement properties [1,2], have become a research hotspot for their potential applications in quantum communication, sensing, and computing [3,4]. Point defects, such as intrinsic vacancies, impurity atoms, and complexes, are promising qubit candidates [5,6], with the NV-center in diamond standing out for its optical addressability and long coherence times at room temperature [7-10]. However, the precise fabrication, localization, and integration of color centers [11], especially in diamond, are experimentally challenging, severely limiting their scalability and, in turn, hindering the development of quantum computing. There is an urgent need for color centers that can be precisely fabricated and localized, along with integration-friendly solutions.

Defect formation and manipulation in two-dimensional (2D) materials are simpler than in three-dimensional (3D) cases, with their electronic states easier to probe and tune. Furthermore, 2D materials can be more easily integrated with other materials. Therefore, many studies have been focused on the potential applications of 2D materials in fields such as quantum computing [12-14]. It has reported that the wide-gap h-BN monolayer exhibit high emission rates and relatively strong zero phonon lines as single-photon emitters (SPEs), and it is easy to be integrated with other optical components [15]. A family of $M_X$ defects in transition metal dichalcogenides monolayer has been proven to be promising candidates for quantum network applications [16]. Especially, the $C^0_S$ in $WS_2$ monolayer has been proven to be a promising qubit candidate for its spin can be optically initialized and read out, and coherent states can be prepared by microwave excitation [17]. However, in these 2D materials, both naturally occurring and artificially induced defects exhibit significant randomness, making it difficult to introduce and precisely locate impurity energy levels from foreign atoms.

The atomic configuration of graphitic carbon nitride (g-$C_3N_4$) [18,19] possesses intrinsic and periodically distributed defects with approximate $C_{2V}$ local symmetry, which are suitable for trapping light atoms like B and C that prefer 3-corrdinated sp² configuration. These defects have the potential to form color centers and offer advantages such as ease of preparation, localization, and integration. Moreover, in terms of electronic property, g-$C_3N_4$ is also a semiconductor with a large band gap of 3.79 [20]. In this work, we systematically evaluated the structures, stabilities and color center properties of B/C-doped g-$C_3N_4$ based on first-principles methods. The vacancy sites are confirmed as the most stable adsorption positions, and once atoms are adsorbed, they are protected by diffusion barriers from thermal diffusions. These B/C-doped defects exhibit optically allowed transition levels, with fluorescence wavelengths in the mid-infrared range, making them form stable qubits for initialization and readout. The Zero-field splitting parameter and hyperfine tensor are provided as potential fingerprints for EPR experiments. These results demonstrate that g-$C_3N_4$ with natural intrinsic vacancies is highly suitable for trapping B/C atom to form stable color centers as qubits.

The main calculations in our work are based on density functional theory (DFT)-related methods and software [21-24]. Specific details can be found in the first section of the supplementary materials. As shown in Fig. 1 (a), the ground state configuration of g-$C_3N_4$ possesses Pca21 space (No. 29) symmetry as predicted in previous work by RG$^2$ code [20,25,26]. It was confirmed as a dynamically stable semiconductor and we consider it as the host material in our present work. After optimization, the lattice constants are a = 9.06 Å, b = 7.98 Å and c = 25 Å, respectively. The optimized C-N bonds and bond angles in g-$C_3N_4$ (Fig. S1 (a)) range from 1.32 Å to 1.42 Å and 114.47° to 126.59°, respectively. The HSE-based electronic band structures further confirmed that g-$C_3N_4$ is a large-gap semiconductor with a direct band gap of 3.78 eV as shown in Fig. 1 (b). These results consistent with previous calculations [20]. As the simulated STM image shown in Fig. 1 (a) and Fig. S2, the regular natural triangular vacancies are distinguishable, which provide highly conducive for doping external C/B atoms to form stable color centers in $C_3N_4$. Especially, the vacancies doped by B/C atoms are clearly visible, which is highly beneficial for the precise localization of color centers.

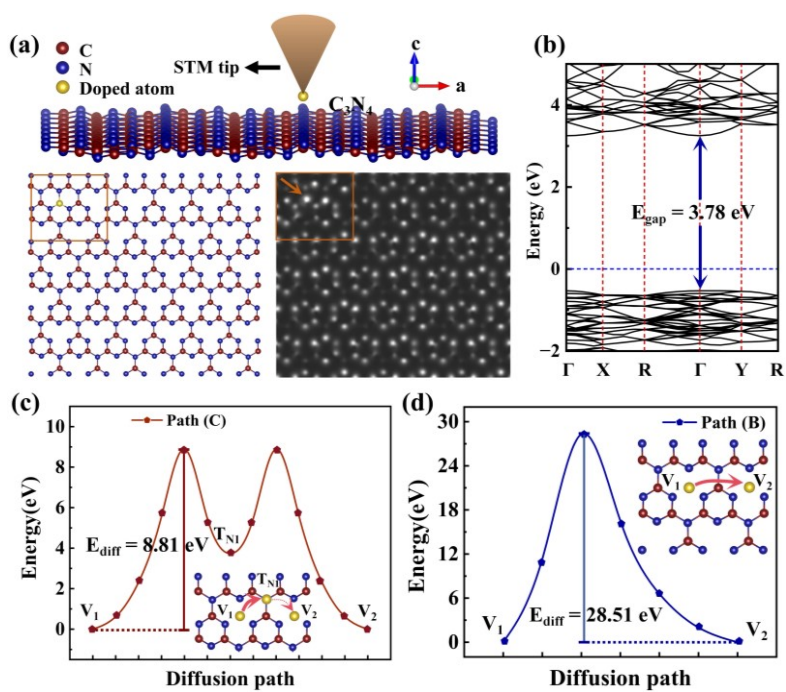

Fig. 1. (a) The atomic structure and corresponding simulated Scanning Tunneling Microscope (STM) image of $C_3N_4$ with a doped C atom. The orange box represents the unit cell. The red, blue, and yellow spheres represent the C, N, and impurity atoms (C/B), respectively. (b) HSE energy band of the $C_3N_4$ supercell. The diffusion paths and the diffusion energy barriers for (c) C and (d) B on the $C_3N_4$ surface.

To assess the capacity of these intrinsic vacancies for trapping doped B/C atoms, five distinct sites are considered as shown in Fig. S1 and Table S1. The results suggest that V-site is the most stable position for trapping both C and B atoms, with adsorption energies of -1.05 eV and -0.99 eV, respectively. To assess whether these adsorption sites can firmly capture B/C atoms to form stable color centers, the diffusion barriers between adjacent sites are calculated. As the shown in Fig. 1 (c) and (d), the diffusion barriers of C and B between adjacent V-sites are different. The former needs to pass through the $T_{N1}$ site, while the latter passes through the bond bridge site between the six-membered rings. The barrier heights along these two paths are denoted as 8.81 eV and 28.51 eV, respectively, which are large enough to hinder the free diffusion of atoms, and also beneficial for the realization of qubits with strong coherence.

The C and N atoms in g-$C_3N_4$ can form perfect bonding relationships and electronic occupation and leading the system strong covalent bond coupling and a wide band gap semiconductor. When foreign B and C atoms are adsorbed into the vacancies, their local geometric and chemical environments are significantly different from the C and N atoms in the material itself, giving them the potential to appear in the middle of the intrinsic band gap. We hope for this, but it still needs to determine where the energy levels of foreign B/C atoms. As shown in Fig. 2 and Fig. S3, they do appear within the original band gap as the highest occupied and lowest unoccupied defect states, which are labeled according to the irreducible representations of $C_{2V}$ point group. For the C with four valence electrons, there are four states in the band gap are occupied, leading to a spin triplet ground state with S = 1, which is similar to the NV⁻ color centers in diamond. These states include three spin-up states ($1b_1^\uparrow$, $2b_1^\uparrow$, $a_2^\uparrow$) and one spin-down state ($1b_1^\downarrow$). The unoccupied spin-down states $2b_1^\downarrow$ appear near the bottom of the conduction band, and another unoccupied spin-down states $a_2^\downarrow$ is inside the conduction band. Thus, the $C_V$ defect has the potential to be used as a qubit, as an electron can be excited from $1b_1^\downarrow$ to $2b_1^\downarrow$, as shown in Fig. 2 (a). For B with 3 valance electrons, there are three occupied states ($a_2^\uparrow$, $1b_1^\uparrow$, $2b_1^\uparrow$) and one unoccupied state ($a_2^\downarrow$) introduced into the band gap, resulting in a net spin of $S = 3/2$ and a local magnetic moment of 3 $\mu_B$. Since the spin-up states are all occupied and all the spin-down states are unoccupied, there is no allowed optical transition in $B_V$ defect. Consequently, the neutral $B_V$ defects cannot be used as a spin qubit in its natural state.

However, whether it has the potential to be used as qubits in the charged state is also an area for further research.

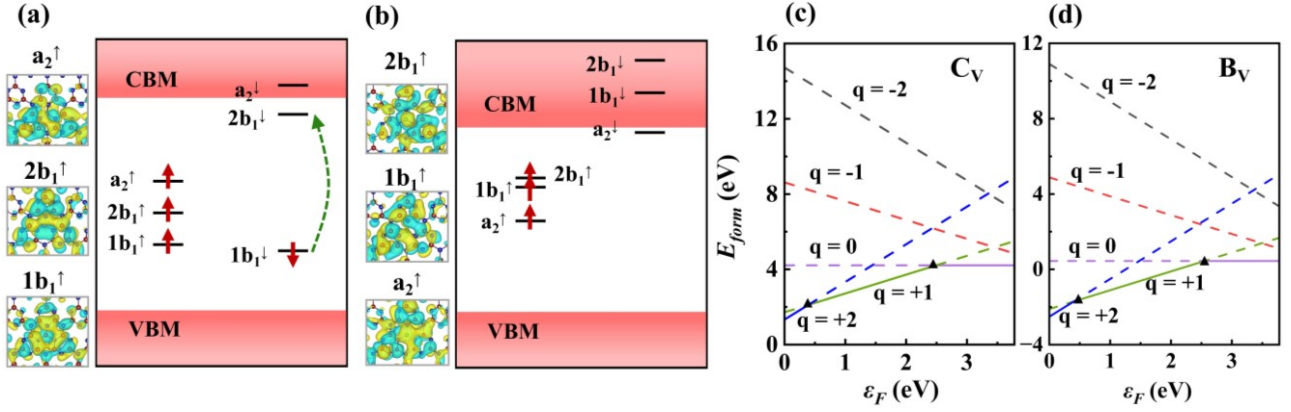

Fig. 2. Schematic diagram of the defect states in the gap of (a) the $C_V$ center ($S = 1$) and (b) the $B_V$ center ($S = 3/2$), along with the real space wavefunction corresponding to each defect state. The red arrow indicates an occupied state, and the green arrow indicates the electron transition path. Yellow and blue correspond to positive and negative components of each real space wave function, respectively. The isosurface value is 3.8% of the maximum value. The formation energy of (c) $C_V$ (N-poor) and (d) $B_v$ defect centers in charge states $q$ as a function of fermi energy in $C_3N_4$. $\varepsilon_F$ represent the fermi energy of pristine $C_3N_4$.

We then pay our attentions to investigate whether the $C_V$ and $B_V$ defects in different charge states exhibit the electronic transition properties required for qubit applications. The defect formation energy is used to evaluate whether a defect can stably exist in a host material [25], with a further charge correction [11,26] as explained in the method part in the supplementary file. The formation energies of the $C_V$ and $B_v$ defects in charge states $q$ as a function of fermi energy are shown in Fig. 2 (c) and (d). $C_V^{+2}$ and $B_V^{+2}$ have the lowest formation energy when $\varepsilon_F$ is between 0 eV to 0.39 eV, while $C_V^{+1}$ and $B_V^{+1}$ have the lowest formation energy when $\varepsilon_F$ is between 0.39 eV and 2.49 eV, respectively. When the formation energy exceeds 2.49 eV, $C_V^0$ and $B_V^0$ become stable. In addition, Fig. 2 (c) shows the formation energy of Cv defect under N-poor condition, and the result under N-rich condition is shown in Fig. S4. The charge transitions of the stable defects in the N-rich condition are the same as those in the N-poor condition, but the defects in N-rich condition are more energetically favorable. Furthermore, the formation energy of Bv defect is lower than that of $C_V$ defect, and it is easier to form.

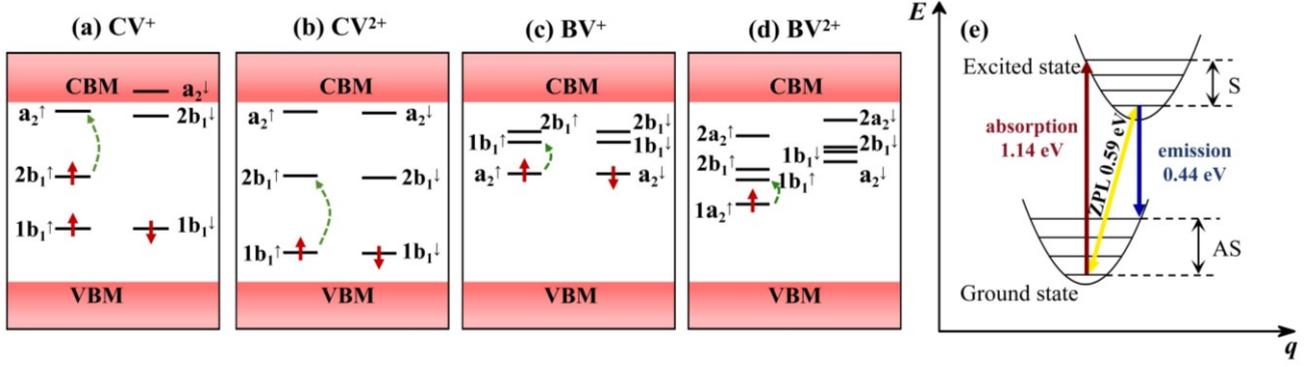

Fig. 3 Schematic diagram of the defect energy levels in the gap of the (a) $C_V^+$ center ($S = 1/2$), (b) $C_V^{2+}$ center ($S = 0$), (c) $B_V^+$ center ($S = 0$), and (d) $B_V^{2+}$ center ($S = 1/2$). (e) The configuration coordinate diagram of $C_V^{2+}$ center. E and q represent the total energy and the configuration coordinate, respectively. The red, blue, and the yellow arrows represent vertical absorption, vertical emission, and the zero-phonon line (ZPL), respectively. S and AS stand for Stokes shift and anti-Stokes shift, respectively.

To investigate the possible excitations defects levels of $C_V$ and $B_V$ under various charge states, the energy level structures of $C_V^+$, $C_V^{2+}$, $C_V^-$, $C_V^{2-}$, $B_V^+$, $B_V^{2+}$, $B_V^-$ and $B_V^{2-}$ have been calculated as shown in Fig. 3 (a)-(d) and Fig. S5. For $C_V^+$ with one electron ionized from the highest occupied state $a_2^\uparrow$, it will handle the spin state of $S=1/2$ and possess a local magnetic moment of 1 $\mu_B$, allowing electrons to be excited from $2b_1^\uparrow$ to $a_2^\uparrow$. Similarly, in the $C_V^{2+}$ with two electrons ionized out of the $a_2^\uparrow$ and $2b_1^\uparrow$, the possible excitation of electron is from the occupied $1b_1^\uparrow/1b_1^\downarrow$ to the unoccupied $2b_1^\uparrow/2b_1^\downarrow$. In the case of $B_V^+$, two electrons in the originally occupied states $1b_1^\uparrow$ and $2b_1^\uparrow$ are ionized, with one is transferred to the adjusted $a_2^\downarrow$ and the other is absent from the material, resulting in the possibility of electron excitation from $a_2^\uparrow$ to $1b_1^\uparrow$ or $a_2^\downarrow$ to $1b_1^\downarrow$. If both electrons are absent from the material, it will form the $B_V^{2+}$ state with a net spin of 1/2 and local magnetic moment of 1 $\mu_B$, and the possible excitation is $1a_2^\uparrow$ to $1b_1^\uparrow$. As shown in Fig. S4, for the cases of $C_V^-$, $C_V^{2-}$, $B_V^-$ and $B_V^{2-}$ with net spins of $S = 1/2$ (1 $\mu_B$), $S = 1$ (2 $\mu_B$), $S = 1/2$ (1 $\mu_B$), respectively, all states in the band gap are filled, and there are no excitations available for use as qubits. These results demonstrate that $C_V^+$, $C_V^{2+}$, $B_V^+$, and $B_V^{2+}$ defect centers are potential qubit.

We then turn our attentions to the optical excited transition properties of the above $C_V^0$, $C_V^{1+}$, $C_V^{2+}$, $B_V^{1+}$ and $B_V^{2+}$ defects. Their zero-phonon line (ZPL), optical absorption/excitation, Stokes shift, and anti-Stokes shift are calculated as shown in configuration coordinate diagram in Fig. 3 (e). Here, only the optical transitions between the ground state and the lowest energy excited state of the defects are considered as indicated in Table I. The ZPLs of $C_V^0$, $C_V^{1+}$, and $C_V^{2+}$ are 0.46 eV, 0.28 eV, and 0.59 eV, respectively. The corresponding optical wavelengths are 2700 nm, 4440 nm, and 2100 nm, respectively, all of which fall within the mid-infrared range. The ZPLs of $B_V^{1+}$ and $B_V^{2+}$ are 0.19 eV and 0.29 eV, respectively, corresponding to optical wavelengths of 6540 nm and 4280 nm in the mid-infrared range. Their low-energy optical transitions, in conjunction with the charge-state tunability, may present advantages in designing specialized optoelectronic devices.

Table I Calculated optical absorption/emission, zero-phonon line (ZPL), Stokes shift (S), anti-Stokes shift (AS) corresponding to the transition levels of $C_V^0$, $C_V^{1+}$, $C_V^{2+}$, $B_V^{1+}$, and $B_V^{2+}$ in $C_3N_4$. The unit of the value is eV.

| Defect, Transition | Absorption | ZPL | Emission | S | AS |
| --- | --- | --- | --- | --- | --- |
| $C_V^0$, $1b_1^\downarrow/2b_1^\downarrow$ | 0.92 | 0.46 | 0.27 | 0.46 | 0.19 |
| $C_V^{1+}$, $2b_1^\uparrow/a_2^\uparrow$ | 0.29 | 0.28 | 0.01 | 0.01 | 0.17 |
| $C_V^{2+}$, $1b_1^\uparrow/2b_1^\uparrow$ | 1.14 | 0.59 | 0.44 | 0.55 | 0.15 |
| $B_V^{1+}$, $a_2^\uparrow/1b_1^\uparrow$ | 0.42 | 0.19 | 0.05 | 0.23 | 0.14 |
| $B_V^{1+}$, $1a_2^\uparrow/2b_1^\uparrow$ | 0.35 | 0.29 | 0.05 | 0.06 | 0.24 |

Finally, it is also need to consider the electron paramagnetic resonance (EPR) properties of these defect states, which are typically described by a spin Hamiltonian [27] (see details in the method part of the supplementary materials) that incorporates the zero-field splitting (ZFS) D tensor and the hyperfine interaction parameter A [28]. In the first step, the ground state ZFS D tensors caused by spin-spin dipole interaction of defects ($C_V^0$, $B_V^0$ and $B_V^{-1}$) with $S \geq 1$ are calculated. The value of axial tensor $D$ and rhombic tensor $E$ are given by $D = 3D_{ZZ}/2, E = (D_{XX} - D_{YY})/2$. For $C_V^0$ defect (S=1), the ground state ZFS components are D = -268.57 MHz and E = 29.74 MHz, these values are D = 74.31 MHz and E = 21.24 MHz for $B_V^0$ defect (S=2/3), as well as D = 421.43 MHz and E = -53.59 MHz for $B_V^{-1}$ defect (S=1), respectively. It can be notice that the ZFS parameter D of $C_V^0$, $B_V^0$ are in the very high frequency (VHF) range, while that of $B_V^{-1}$ is in the range of ultra-high frequency (UHF).

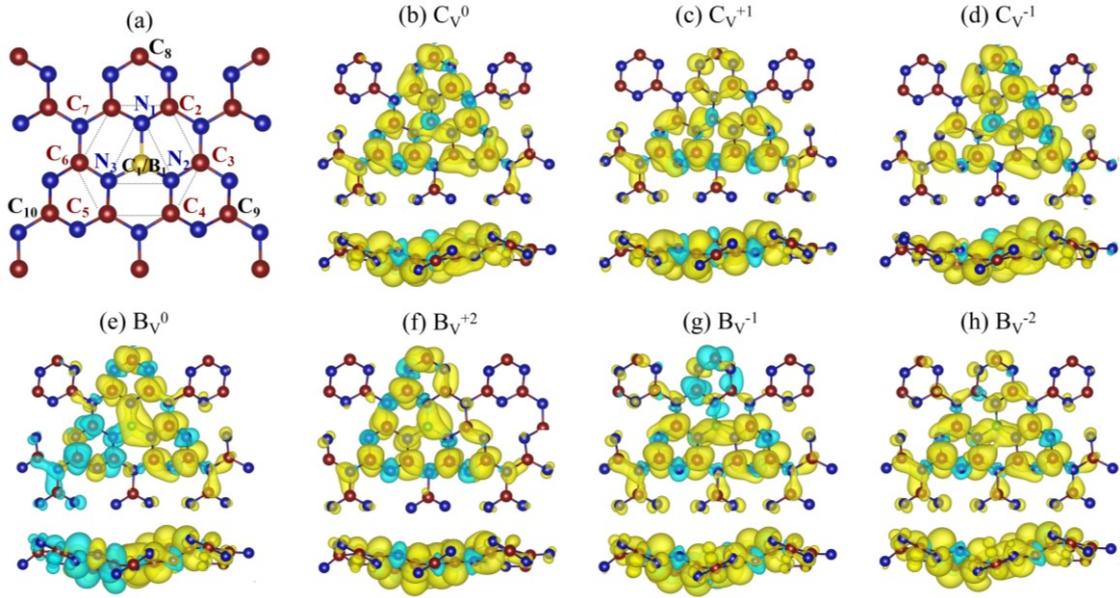

Fig. 4 (a) Local atomic structure of defect. Spin charge density difference of the (b) $C_V^0$, (c) $C_V^{+1}$, (d) $C_V^{-1}$, (e) $B_V^{+1}$, (f) $B_V^{+2}$, (g) $B_V^{-1}$, and (h) $B_V^{-2}$ defect. The $C_1$, $B_1$ represent vacancy doped C or B atoms, $N_1$ - $N_3$ represents the first neighbor N atom, $C_2$ - $C_7$ represents the second neighbor C atom, and $C_8$ - $C_{10}$ represents the third neighbor C atom. The isosurface value is 1.2% for $C_V$ and 1.4% for $B_V$ of the maximum value.

The nuclear spin of $^{13}C$, $^{14}N$ and $^{10}B$ are 1/2, 1 and 3, respectively. There is an intention to investigate the hyperfine parameters C/B doped $C_3N_4$. The hyperfine interaction parameters of $C_V$ and $B_V$ in the charge states with net spin are calculated as shown in Table II. The principal value A is related to the distance from the impurity atom (C/B), thus the position of vacancy atom, nearest neighbor atom and next neighbor atom are shown in Fig. 4 (a). For $C_V$ and $B_V$ defects in all charge states, the impurity atom ($C_1$), and its second ($C_2$-$C_7$), and third ($C_8$-$C_{10}$) nearest neighbors show the largest principal values of hyperfine parameters, followed by the first nearest neighbor N atom ($N_1$ - $N_3$), as shown in Table II. $C_V^{+1}$ and $B_V^0$ have large hyperfine parameter values among the defects, reaching 102.22 MHz and 132.77 MHz in Azz, respectively. Due to the symmetry weakening caused by the fluctuations, the distances from the vacancy to the various nearest neighbor positions around it are no longer the same. Thus, their hyperfine parameters are slightly different, as shown in Table SII to Table SVIII for details. Additionally, the spin charge densities are also investigated as the results shown in Fig. 4 (b)-(h). Clearly, they are concentrated on $C_1$, $C_2$-$C_7$ and $C_8$-$C_{10}$, while the distributions on N atoms are always small, consistent with the results of hyperfine parameter values. Therefore, EPR can be measured using these values to provide accurate identification of defects.

Table II Hyperfine interaction tensor principal values ($A_{XX}$, $A_{YY}$, and $A_{ZZ}$) for vacancy atoms ($^{13}C/^{11}B$), neighbor $^{13}C$ atoms ($C_2$-$C_7$, $C_8$-$C_{10}$) and nearest neighbor $^{14}N$ atoms ($N_1$-$N_3$), electron spin of the $C_V$ and $B_V$ defect. Take the maximum value atom among the equivalent position atoms.

| Defect, spin | Atoms | Axx (MHz) | Ayy (MHz) | Azz (MHz) |
|---|---|---|---|---|
| $C_V^0$, 1 | $C_1$ | 24.60 | 16.17 | 28.12 |
| | $C_2$-$C_7$ | 16.36 | 16.05 | 62.27 |
| | $C_8$-$C_{10}$ | 15.34 | 14.70 | 55.46 |
| | $N_1$-$N_3$ | 1.48 | 1.41 | 2.57 |
| $C_V^{+1}$, 1/2 | $C_1$ | 22.42 | 16.78 | 29.13 |
| | $C_2$-$C_7$ | 24.53 | 23.97 | 102.22 |
| | $C_8$-$C_{10}$ | 19.80 | 18.66 | 66.52 |
| | $N_1$-$N_3$ | 1.37 | 1.26 | 3.16 |
| $C_V^{-1}$, 1/2 | $C_1$ | 13.62 | 3.50 | 15.68 |
| | $C_2$-$C_7$ | 16.44 | 16.07 | 72.24 |
| | $C_8$-$C_{10}$ | 17.89 | 17.17 | 62.58 |
| | $N_1$-$N_3$ | 1.13 | 1.09 | 2.94 |
| $B_V^0$, 3/2 | $B_1$ | -8.72 | 7.96 | -15.74 |
| | $C_2$-$C_7$ | 24.63 | 23.54 | 95.21 |
| | $C_8$-$C_{10}$ | 37.32 | 35.91 | 132.77 |
| | $N_1$-$N_3$ | 3.61 | -3.54 | -15.00 |
| $B_V^{+2}$, 1/2 | $B_1$ | 33.80 | 33.00 | 44.41 |
| | $C_2$-$C_7$ | 23.90 | 23.63 | 88.03 |
| | $C_8$-$C_{10}$ | 15.34 | 14.56 | 48.32 |
| | $N_1$-$N_3$ | 12.04 | 11.84 | 28.57 |
| $B_V^{-1}$, 1 | $B_1$ | 12.44 | 10.48 | 19.42 |
| | $C_2$-$C_7$ | 18.99 | 18.13 | 65.49 |
| | $C_8$-$C_{10}$ | 29.16 | 28.29 | 103.31 |
| | $N_1$-$N_3$ | 2.97 | 2.74 | 13.67 |
| $B_V^{-2}$, 1/2 | $B_1$ | 8.96 | 8.52 | 14.47 |
| | $C_2$-$C_7$ | 18.58 | 17.93 | 66.80 |
| | $C_8$-$C_{10}$ | 24.65 | 23.97 | 89.10 |
| | $N_1$-$N_3$ | 2.47 | 2.35 | 12.88 |

In summary, we demonstrate that g-$C_3N_4$ with natural intrinsic vacancies is highly suitable for trapping B/C atom to form stable color centers as qubits, which may fill the gap of g-$C_3N_4$ in quantum information application. Specifically, the natural triangular vacancies show strong adsorption energy for doped C/B, which makes C/B protected by high diffusion energy barriers and is also conducive to the realization of strongly coherent qubits. For the neutral $C_V$ defect, an electron can be excited from $1b_1^\downarrow$ to $2b_1^\downarrow$, while there is no allowed optical transition in BV defect. In addition, the defect formation energies indicate that the most stable charge states are $C_V^{+2}/B_V^{+2}$, $C_V^{+1}/B_V^{+1}$ and $C_V^0/B_V^0$ in turn. The net spin and possible transition defects levels of $C_V$ defect and $B_V$ defect can be tuned by charge states, among which $C_V^+$, $C_V^{2+}$, $B_V^+$, and $B_V^{2+}$ are the potential qubits. The ZPLs of $C_V$, $C_V^+$, $C_V^{2+}$, $B_V^+$, and $B_V^{2+}$ fall within the mid-infrared fluorescence range which is conducive to precise initialization. Furthermore, the zero-field splitting and hyperfine coupling parameters of the defects are provided for serving as guiding parameters of the EPR spectrum. This work provides new ideas for the application of 2D materials with intrinsic vacancies as qubits hosts.

**Acknowledgement**

This work is supported by the National Natural Science Foundation of China (Grant No. 12374046), the Shanghai Science and Technology Innovation Action Plan (Grant No. 24LZ1400800), the Program of Changjiang Scholars and Innovative Research Team in University (Grant No. IRT-17R91), the Scientific Research Foundation of Hunan Provincial Education Department (Grant No. 23B0182), the Natural Science Foundation of Hunan Province (Grant No. 2022JJ40420), the Postgraduate Scientific Research Innovation Project of Hunan Province (Grant No. LXBZZ2024116).

**Reference**

[1]  A. Chatterjee, P. Stevenson, S. De Franceschi, A. Morello, N. P. de Leon, and F. Kuemmeth, Nat. Rev. Phys. **3**, 157 (2021).
[2]  M. Körber, O. Morin, S. Langenfeld, A. Neuzner, S. Ritter, and G. Rempe, Nat. Photonics **12**, 18 (2017).
[3]  P. Tian, L. Tang, K. S. Teng, and S. P. Lau, Mater. Today Chem. **10**, 221 (2018).
[4]  Y. Lu, A. Sigov, L. Ratkin, L. A. Ivanov, and M. Zuo, J. Ind. Inf. Integration **35** (2023).

[5]  P. M. Koenraad and M. E. Flatte, Nat. Mater. **10**, 91 (2011).
[6]  C. Bradac, W. Gao, J. Forneris, M. E. Trusheim, and I. Aharonovich, Nat. Commun. **10**, 5625 (2019).
[7]  C. Kurtsiefer., S. Mayer., P. Zarda., and H. Weinfurter, Phys. Rev. Lett. **85**, 290 (2000).
[8]  A. Gali, E. Janzen, P. Deak, G. Kresse, and E. Kaxiras, Phys. Rev. Lett. **103**, 186404 (2009).
[9]  F. M. Hossain, M. W. Doherty, H. F. Wilson, and L. C. Hollenberg, Phys. Rev. Lett. **101**, 226403 (2008).
[10] F. Shi *et al.*, Phys. Rev. Lett. **105**, 040504 (2010).
[11] H. J. von Bardeleben, S. A. Zargaleh, J. L. Cantin, W. B. Gao, T. Biktagirov, and U. Gerstmann, Phys. Rev. Mater. **3**, 124605 (2019).
[12] Z. Benedek, R. Babar, Á. Ganyecz, T. Szilvási, Ö. Legeza, G. Barcza, and V. Ivády, npj Comput. Mater. **9**, 187 (2023).
[13] P. V. Bakharev *et al.*, Nat. Nanotechnol. **15**, 59 (2020).
[14] S. Gupta, J. H. Yang, and B. I. Yakobson, Nano Lett. **19**, 408 (2019).
[15] K. Li, T. J. Smart, and Y. Ping, Phys. Rev. Mater. **6**, 042201 (2022).
[16] Y. Lee, Y. Hu, X. Lang, D. Kim, K. Li, Y. Ping, K. C. Fu, and K. Cho, Nat. Commun. **13**, 7501 (2022).
[17] S. Li, G. Thiering, P. Udvarhelyi, V. Ivady, and A. Gali, Nat. Commun. **13**, 1210 (2022).
[18] S. Patnaik, D. P. Sahoo, and K. Parida, Carbon **172**, 682 (2021).
[19] Y. Zheng, J. Liu, J. Liang, M. Jaroniec, and S. Z. Qiao, Energy Environ. Sci. **5**, 6717 (2012).
[20] L. zhao, X. Shi, J. Li, T. Ouyang, C. Zhang, C. Tang, C. He, and J. Zhong, Physica E Low Dimens. Syst. Nanostruct. **128**, 114535 (2021).
[21] G. Kresse and J. Hafner, Phys. Rev. B Condens. Matter **47**, 558 (1993).
[22] G. Kresse and J. Furthmüller, Phys. Rev. B **54**, 11169 (1996).
[23] J. P. Perdew., K. Burke., and M. Ernzerhof., Phys. Rev. Lett. **77**, 3865 (1996).
[24] G. Henkelman, B. P. Uberuaga, and H. Jónsson, J. Chem. Phys. **113**, 9901 (2000).
[25] J. B. Varley, A. Janotti, and C. G. Van de Walle, Phys. Rev. B **93**, 161201 (2016).
[26] C. Freysoldt and J. Neugebauer, Phys. Rev. B **97**, 205425 (2018).
[27] S. Xu, M. Liu, T. Xie, Z. Zhao, Q. Shi, P. Yu, C.-K. Duan, F. Shi, and J. Du, Phys. Rev. B **107**, 140101 (2023).
[28] I. Takács and V. Ivády, Commun. Phys. **7**, 178 (2024).